%% file: wLFitting_paper.tex
\documentclass[useAMS,usenatbib]{mn2e}
\usepackage{graphicx}
\usepackage{amssymb, amsmath}
\usepackage{times}
\usepackage{color}
\usepackage{lscape}
\usepackage{float}
\usepackage{array}
\newcommand{\wL}{\omega_{\ell}\left(r_{c}\right)}
\newcommand{\Hp}{H\left(\lowercase{z}\right)}
\newcommand{\Da}{D_A\left(\lowercase{z}\right)}

\newcommand{\comment}[1]{\relax}		    % remove all comments

\begin{document}
\title[BAO Fitting with $\wL$] {Robust New Statistic for fitting the Baryon Acoustic Feature}%{Fitting the Baryon Acoustic Feature with the $\wL$ statistic}

\input{authors}

\date{\today} 
\pagerange{\pageref{firstpage}--\pageref{lastpage}} \pubyear{2014}
\maketitle
\label{firstpage}

\begin{abstract}

\input{abstract}

\end{abstract}


\maketitle

%Section1
\section{Introduction}\label{sec:intro}
\input{intro} %TBC

%Section 2
\section{Data and Mocks}\label{sec:data}
\input{data} %TBC

%Section 3
\section{Background and Theory}\label{sec:theory}
\input{theory}

%Section 4
\section{Analysis}\label{sec:analysis}
\input{analysis}

%Section 5
\section{Fitting Results}\label{sec:results}
\input{results} 

%Section 6
\section{Advantages of $\wL$}\label{sec:advantages}
\input{advantages}

%Section 7
\section{Conclusions}\label{sec:conc}
\input{conclusion}

\section{Acknowledgements}\label{sec:acknow}
\input{acknowledgement}

\input{bibliography}
\end{document}

%% file: authors.tex
\author[Osumi et al.]{\parbox{\textwidth}{\Large
Keisuke Osumi$^{1}$,
Shirley Ho$^{1}$,
Daniel J. Eisenstein$^{2}$,
Mariana Vargas-Maga\~na$^{1}$,
 } \vspace*{4pt} \\ 
 \scriptsize $^{1}$ Department of Physics, Carnegie Mellon University, 5000 Forbes Avenue, Pittsburgh, PA 15213, USA\vspace*{-2pt} \\ 
 \scriptsize $^{2}$ Harvard-Smithsonian Center for Astrophysics, 60 Garden Street, Cambridge MA 02138, USA\vspace*{-2pt} \\ 
}

%% file: abstract.tex
We investigate the utility and robustness of a new statistic, $\wL$, for analyzing Baryon Acoustic Oscillations (BAO). 
We apply $\wL$, introduced in \cite{Xu10}, to mocks and data from the Sloan Digital Sky Survey (SDSS)-III Baryon Oscillation Spectroscopic Survey (BOSS) included in the SDSS Data Release Eleven (DR11). 
We fit the anisotropic clustering using the monopole and quadrupole of the $\wL$ statistic in a manner similar to conventional multipole fitting methods using the correlation function as detailed in \citep{Xu12}. 
To test the performance of the $\wL$ statistic we compare our results to those obtained using the multipoles.
The results are in agreement. 
We also conduct a brief investigation into some of the possible advantages of using the $\wL$ statistic for BAO analysis. 
The $\wL$ analysis matches the stability of the multipoles analysis in response to artificially introduced distortions in the data, without using extra nuisance parameters to improve the fit.
When applied to data with systematics, the $\wL$ statistic again matches the performance of fitting the multipoles without using nuisance parameters.
In all the analyzed circumstances, we find that fitting the $\wL$ statistic removes the requirement for extra nuisance parameters.

%% file: intro.tex
Baryon Acoustic Oscillations (BAO) are a powerful tool to break degeneracies in measurements that rely solely on the cosmic microwave background (CMB).
For example, CMB measurements alone cannot obtain satisfactory constraints on the Hubble constant, $H_0$, and the matter density, $\Omega_m$ \citep{Hin13}.

BAO arise from acoustic oscillations before recombination; their effects are imprinted on the galaxy distribution in the universe \citep{Eis98}.
The scale of these oscillations can be used as a standard ruler to provide distance measurements to various redshifts.
These measurements are sensitive to the value of cosmological parameters and provide independent observations which break CMB degeneracies. 

With the most recent galaxy surveys, it is possible to measure the BAO feature in both the line-of-sight and transverse directions.
This constrains both the Hubble parameter, $\Hp$, and the angular diameter distance, $\Da$, at redshift $z$.
Because the BAO feature is sensitive to cosmology, the feature shifts in the line-of-sight and transverse directions based on the fiducial cosmology used during analysis.
By applying the Alcock-Paczynski test, we can use these measurements to constrain cosmology \citep{AP}.

Over the past few years, several statistics have been proposed for analyzing the BAO feature in galaxy surveys.
In particular, the multipoles fitting method has been applied to data from the SDSS-III BOSS Data Releases to produce consistent constraints on $\Hp$ and $\Da$ \citep{Xu12}. 
In this paper, we focus on a statistic, $\wL$, which may provide several advantages over ones in current use \citep{Xu10}.
We apply the $\wL$ statistic to data and mock catalogs from the BOSS Data Release 11 and provide comparisons to constraints produced by the multipoles method.
We also perform a brief investigation into the possible advantages of working with the $\wL$ statistic. 

The remainder of this paper is structured as follows. 
In Section~\ref{sec:data} we describe the data and mocks used in this paper. 
In Section~\ref{sec:theory} we give a brief introduction of the theory behind the fitting method used. 
In Section~\ref{sec:analysis} we present the details of the fitting model used in our analysis. 
In Section~\ref{sec:results} we present our results from applying the $\wL$ statistic to mock catalogs and data from the SDSS-III Data Release 11 (DR11).
In Section~\ref{sec:advantages} we describe our investigation into the advantages of using the $\wL$ statistic.
Finally, we conclude in Section~\ref{sec:conc}.

%% file: data.tex
\subsection{Data}
\label{subsec:data}

We use data included in Data Release 11 (DR11) of the Sloan Digital Sky Survey (SDSS; \cite {Yor00}).
Using the 2.5m Sloan Telescope \citep{Gun06} at Apache Point Observatory in New Mexico, SDSS I, II \citep{Aba09}, and III \citep{Eis11} used a drift-scanning mosaic CCD camera \citep{Gun98} to image 14,555 square degrees of the sky in five photometric bandpasses (\cite{Fuk96}; \cite{Smi02}; \cite{Doi10}) to a limiting magnitude of $\simeq$ 22.5.
The data was passed through pipelines designed to perform astrometric calibration \citep{Pie03}, photometric reduction \citep{Lup01}, and photometric calibration \citep{Pad08}. 
All of the imaging data was re-processed as part of SDSS Data Release 8 (DR8; \cite{Aih11}.

The Baryon Oscillation Spectroscopic Survey (BOSS) itself is designed to obtain spectra and redshifts for 1.35 million galaxies over a footprint of approximately 10,000 square degrees with a redshift completeness of over 97 percent.
The targets are selected from SDSS DR8 imaging and assigned to tiles of diameter 3$^{\circ}$ using a target density adaptive algorithm \citep{Bla03}.
Redshifts are derived from spectra \citep{Bol12} obtained using the doubled-armed BOSS spectrographs \citep{Sme13}. 
For a summery of the survey, one may consult \cite{Eis11}. 
\cite{Daw12} also provides a full description.

\subsection{Simulations}
\label{subsec:sims}

We use 600 SDSS III-BOSS DR11 PTHalos mock galaxy catalogs with the same angular and radial masks as the survey data to compute the sample covariance matrices used in our analysis.
These mock catalogs, provided by \cite{Man13}, are generated at $z=0.55$ in boxes of side length $2400 h^{-1}$Mpc with $1280^3$ dark matter particles. 

When generating the mocks, \cite{Man13} first used second-order Lagrangian perturbation theory (2LPT) to create a matter density field. 
Then, the halos were identified using a friends of friends halo finder.
Masses and linking lengths were calibrated to N-body simulations, and the halos were populated using a Halo Occupation Distribution calibrated  to match the observed clustering in CMASS catalog \citep{Whi11}. 

%% file: theory.tex
\subsection{Background and Parameterization}
\label{ssec:background}

There are two main effects that contribute to anisotopic clustering: redshift-space distortions and the adoption of an incorrect cosmology when calculating galaxy separations. 
Redshift-space distortions are generated by the peculiar motions of galaxies within clusters (Finger-of-God effect) and by larger scale flows of galaxies into overdense regions \citep{Kai87}.
These effects distort our measurements of line of sight separation between galaxies and affect the shape of the correlation function. 

Adopting an incorrect cosmology distorts our measurements of galaxy separations in both the line-of-sight and transverse directions. 
Line of sight separations are computed using the Hubble parameter $\Hp$ while transverse separations are computed using the angular diameter distance $\Da$. 
However, both $\Hp$ and $\Da$ are dependent on cosmology. 
As a result, adopting an incorrect cosmology gives a different clustering signal and BAO signal in the line-of-sight direction versus the transverse direction \citep{Xu12}. 
In addition to these anisotropic effects, the BAO feature can also be shifted isotropically as a result of assuming an incorrect cosmology. 
By applying the Alcock-Paczynski test, we can use these shifts to constrain our cosmology \citep{AP}. 

To measure the anisotropy we construct a clustering model that parameterizes the BAO shifts. 
The parameters can then be measured by fitting the model to our data. 
As in \cite{Xu12}, we parameterize the isotropic shift using $\alpha$ and the anisotropic signal using $\epsilon$. 
These parameters are defined by:
\begin{equation}
\alpha = \frac{D_V\left(z\right)/r_s}{D_{V,fid}\left(z\right)/r_{s,fid}}
\end{equation}
\begin{equation}
1 + \epsilon = {\left[\frac{H_{fid}\left(z\right)}{\Hp}\frac{D_{A,fid}\left(z\right)}{\Da}\right]}^{1/3}
\end{equation}
%Remember to define D_V in the intro!
Here, $r_s$ denotes the sound horizon or the BAO scale. The subscript $fid$ denotes the fiducial cosmology (see Section~\ref{ssec:fidcosmology}) which we assume when making our measurements.
$D_V\left(z\right)$ is the spherically averaged distance to redshift z. 
It is defined by:
\begin{equation}
D_V\left(z\right)=\left[\left(1+z\right)^2D_A^2\left(z\right)\frac{cz}{H\left(z\right)}\right]^{1/3}
\end{equation}

Note that if there is no isotropic shift, then $\alpha = 1$ and if there is no anisotropic warping, $\epsilon = 0$. 

Alternatively, we can also parameterize using the shift parallel to the line-of-sight, $\alpha_{||}$, and the shift perpendicular to the line-of-sight , $\alpha_{\perp}$:
\begin{equation}
\alpha_{||} =\frac{\Da r_{s, fid}}{D_{A, fid}\left(z\right)r_s}
\end{equation}
\begin{equation}
\alpha_{\perp} =\frac{H_{fid}\left(z\right)r_{s,fid}}{\Hp r_s}
\end{equation}

The relations between the two parameterizations are given by:
\begin{equation}
\alpha=\alpha_\perp^{2/3}\alpha_{||}^{1/3}
\end{equation}
\begin{equation}
1+\epsilon=\left (\frac{\alpha_{||}}{\alpha_{\perp}}\right)^{1/3}
\end{equation}

\subsection{Clustering Estimators}
\label{ssec:cestimators}

Most clustering estimators for measuring the BAO feature require either the computation of the 2D correlation function, $\xi\left(s,\mu\right)$, or the power spectrum, $P\left(k\right)$ \citep{And14}.
The two estimators we investigate here, Multipoles \citep{Xu12} and the $\wL$ statistic (\cite{Xu10}; \cite{Bla11}), are based on analysis of the 2D correlation function. 
However, working with the full 2D correlation function is impractical as it requires a far larger quantity of mock catalogs than are currently available.
To avoid this problem, we perform our analysis by compressing the correlation function into a small number of angular moments.  

\subsubsection{Multipoles}
\label{sssec:multipoles}

Since the primary focus of this paper is the $\wL$ statistic, we will only present a summary of the multipole analysis. 
The interested reader may consult \cite{Xu12} for a more detailed discussion. 

First we define our coordinates:
\begin{equation}
r^2 = r_{||}^2 + r_{\perp}^2
\end{equation}
\begin{equation}
\mu = cos\theta = \frac{r_{||}}{r}
\end{equation}
Here, $r$ is the separation between two galaxies, $r_{||}$ is the separation in the line-of-sight direction, and $r_{\perp}$ is the separation in the transverse direction. $\theta$ is the angle between a galaxy pair and the line-of-sight. 

Then, $\alpha$ and $\epsilon$ are defined by:
\begin{equation}
r_{||}' = \alpha\left(1+\epsilon\right)^2r_{||}
\end{equation}
\begin{equation}
r_{\perp}' = \alpha\left(1+\epsilon\right)^{-1}r_{\perp}
\end{equation}
where the primed coordinates denote the true cosmology space and the unprimed the fiducial cosmology space. 

The multipole analysis measures the Legendre moments of the 2D correlation function:
\begin{equation}
\xi_{\ell}\left(r\right) = \frac{2\ell + 1}{2} \int_{-1}^{1} d\mu \ \xi\left(r',\mu\right)L_{\ell}\left(\mu\right)
\end{equation}
where $L_{\ell}\left(\mu\right)$ is the $\ell$th Legendre polynomial.

In the multipole analysis, we focus on the monopole and quadrupole since the influence of higher order terms is negligible.
Note that the odd order multipoles are zero due to symmetry.

\subsubsection{The $\wL$ Statistic}
\label{sssec:wLstatistic}

As in \cite{Xu10}, we define $\wL$ as the redshift space correlation function, $\xi_s\left(r, \mu\right)$, convolved with a compact and compensated filter $W_{\ell}\left(r,\mu, r_{c}\right)$ as a function of characteristic scale $r_{c}$:
\begin{equation}
\label{eqn:wLdef}
\begin{aligned}
\wL & = i^{\ell}\int d^3r \ \xi_s\left(r,\mu\right)W_{\ell}\left(r,\mu, r_{c}\right)\\
& = i^{\ell}\int d^3r \ \xi_s\left(r,\mu\right)W_{\ell}\left(r, r_c\right)L_{\ell}\left(\mu\right)\\
& = \frac{4\pi i^{\ell}}{2\ell + 1}\int r^2 \ dr \ \xi_{\ell}\left(r\right)W_{\ell}\left(r,r_{c}\right)
\end{aligned}
\end{equation}
where we have taken advantage of the orthogonality of the Legendre polynomials and set $W_{\ell} \left(r,\mu, r_c \right) = W_{\ell}\left(r, r_c \right) L_{\ell}\left(\mu\right)$. 
%Note that the expected measured $\omega_0\left(r_c\right)$ and $\omega_2\left(r_c\right)$ are generated by applying Eq.~\ref{eqn:wLdef} to Eq.~\ref{eqn:expectedxi0} and Eq.~\ref{eqn:expectedxi2}. 

Following \cite{Pad07} and \cite{Xu10} we define a smooth, low order, compensated filter independently of $\ell$:
\begin{equation}
\label{eqn:filter}
W\left(x\right) = \left(2x\right)^2\left(1-x\right)^2\left(\frac{1}{2}-x\right)\frac{1}{r_c^3}
\end{equation}
where $x = \left(r/r_c\right)^3$. 
%include figure here: plot of filter?

This choice of filter gives the $\wL$ statistic several advantages.
By design, $\wL$ probes a narrow range of scales near the BAO feature and is not sensitive to large scale fluctuations or to poorly measured or modeled large scale modes \citep{Xu10}. 
Further, $\wL$ is blind to the constant $\xi$ monomial, which is the integral constraint term that $\wL$ is designed to avoid. Any other monomials in $\xi$ also produce the same monomial in $\wL$. 

\subsection{Modeling the correlation function}
\label{ssec:modeling}
A model for the correlation function is required in order to perform our fitting analysis with our two estimators. 
We start with a nonlinear power spectrum template:
\begin{equation}\label{eqn:2dPS}
P\left(k,\mu\right) = \left(1+\beta\mu^2\right)^2F\left(k,\mu,\Sigma_s\right)P_{dw}\left(k,\mu,\Sigma_{\perp},\Sigma{||}\right)
\end{equation}
where the $\left(1+\beta\mu^2\right)^2$ term corresponds to the Kaiser model \citep{Kai87} for large scale redshift space distortions.
The $F\left(k,\mu,\Sigma_s\right)$ term is the streaming model for the Finger-of-God effect \citep{Pea94}:
\begin{equation}\label{eqn:FOG}
F\left(k,\mu,\Sigma_s\right)=\frac{1}{1+k^2\mu^2\Sigma_s^2}
\end{equation}
where $\Sigma_s$ is the streaming scale. %mention value of Sigma_s here?
$P_{dw}\left(k,\mu,\Sigma_{\perp},\Sigma{||}\right)$ is the ``De-Wiggled'' power spectrum adopted by \cite{Xu12} and \cite{And13}, among others, as a template for the non-linear power spectrum.
$\Sigma_{\perp}$ and $\Sigma{||}$ serve to parameterize the Gaussian damping of BAO due to nonlinear structure growth \citep{Eis07}.

We produce the multipoles of the correlation function by first decomposing the 2D power spectrum given in Eq.~\ref{eqn:2dPS}:
\begin{equation}
P_{\ell}\left(k\right)=\frac{2\ell+1}{2}\int_{-1}^{1}P\left(k,\mu\right)L_{\ell}\left(\mu\right)d\mu
\end{equation}
We then transform to configuration space:
\begin{equation}\label{eqn:expectedxis}
\xi_{\ell}\left(r\right)=i^{\ell}\int\frac{k^3dlog\left(k\right)}{2\pi^2}P_{\ell}\left(k\right)j_{\ell}\left(kr\right)
\end{equation}
where $j_{\ell}\left(kr\right)$ is the $\ell$th spherical Bessel function. 

\subsection{Covariance Matrix}\label{ssec:covm}
In this work, we apply corrections to the covariance matrix as suggested in \cite{Per13}.
Further, since we estimate the covariance matrix from mock catalogs which parallel BOSS observations by splitting the observations into the Northern Galactic Cap (NGC) and the Southern Galactic Cap (SGC), we must adjust the standard computation for the covariances accordingly.
For a detailed description of our computation, the reader may consult \cite{Var13}.

The $ij$th entry in the covariance matrix, $C$, used in our fitting analysis is computed using the 600 PTHalos mocks:
\begin{equation}
\begin{aligned}
C_{ij} = &\frac{1}{2}[\frac{1}{299}\sum_{m\leq 300}\left(\xi^m_i - \bar{\xi}\right)\left(\xi^m_j - \bar{\xi}\right)\\
&+ \frac{1}{299}\sum_{m>300}\left(\xi^m_i - \bar{\xi}\right)\left(\xi^m_j - \bar{\xi}\right)]
\end{aligned}
\end{equation}
where $\xi = \left[\xi_0,\xi_2\right]$, $\bar{\xi}$ is the average over all the mocks, and $\xi^m$ is the correlation function for the $m$th mock.
The first 300 mocks correspond to the NGC and the rest correspond to the SGC.
When computing $C$ for the $\wL$ statistic, we replace the $\xi_{\ell}$'s with $\wL$'s and carry out the same computation. 

When estimating the covariance matrix from a finite number of mocks we introduce a bias which is corrected by multiplying the inverse covariance matrix by $\left(1-D\right)$.
$D$ is defined by:
\begin{equation}
D = \frac{n_b+1}{n_s-1}
\end{equation}
Here, $n_b$ is the number of bins used in the analysis and $n_s$ is the number of mocks used.

To account for the effective volume of our sample, we scale $D$ by:
\begin{equation}
r = \frac{2V_{overlap}}{V_{NGC} + V_{SGC}}
\end{equation}
For the DR11 sample, we have $r = 0.49$. 

\begin{figure}
\includegraphics[width=\linewidth]{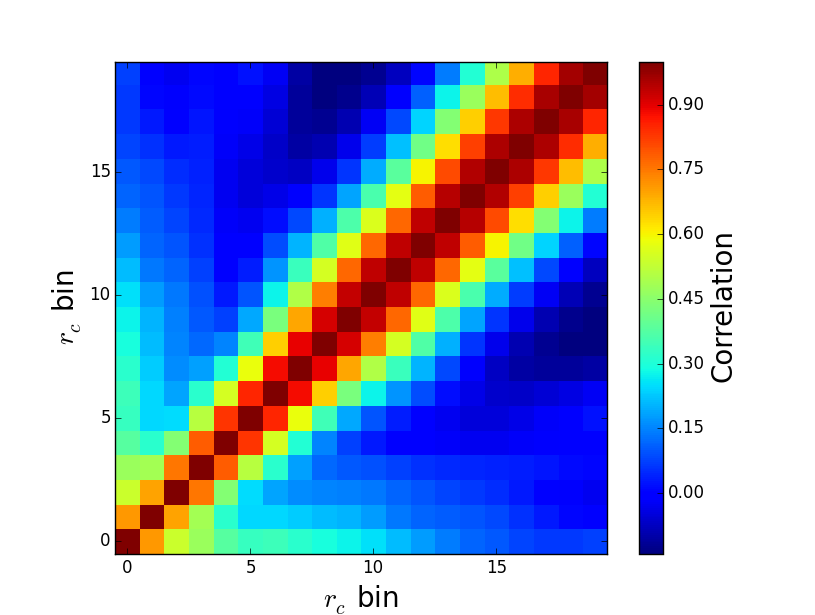}
\includegraphics[width=\linewidth]{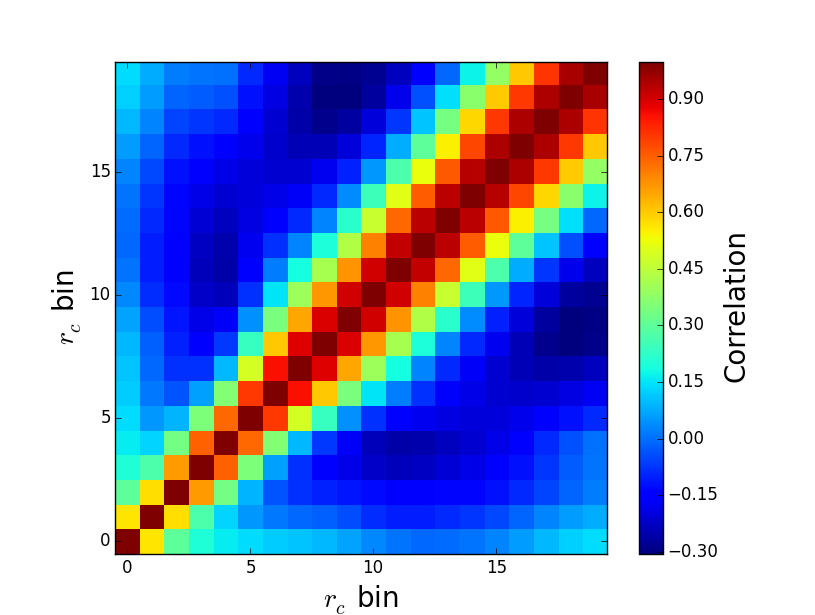}
\caption{The correlation matrix for $\wL$ computed for our analysis. The top figure depicts the covariance in $\omega_{0}\left(r_c\right)$ and the bottom figure depicts the covariance in $\omega_{2}\left(r_c\right)$. Here, the $r_c$ bins run from $46h^{-1}$Mpc to $198h^{-1}$Mpc in bins of size $8h^{-1}$Mpc.}\label{fig:wLredcov}
\end{figure}

The correlation matrix used when fitting the $\wL$ statistic is presented in figure~\ref{fig:wLredcov}.

%% file: analysis.tex
\subsection{Fiducial Cosmology}
\label{ssec:fidcosmology}

Throughout our analysis we assume a fiducial $\Lambda$CDM cosmology with $\Omega_M = 0.274$, $\Omega_b = 0.0457$, $h = 0.7$, and $n_s = 0.95$. 
In this fiducial cosmology, the angular diameter distance to $z=0.57$ is $D_A\left(0.57\right) = 1359.72$ Mpc, the Hubble parameter is $H\left(0.57\right) = 93.56$ kms$^{-1}$Mpc$^{-1}$, and the sound horizon is $r_s = 153.19$ Mpc. 

\subsection{Measuring the Correlation Function and $\wL$}
\label{ssec:measuringxi}

Our correlation functions are computed in radial bins of 8$h^{-1}$Mpc, running from 6$h^{-1}$Mpc to 198$h^{-1}$Mpc, and $\mu$ bins of 0.01 using the Landy-Szalay estimator \citep{Lan93}:
\begin{equation}
\xi\left(r,\mu\right) \cong \frac{DD\left(r,\mu\right)-2DR\left(r,\mu\right)+RR\left(r,\mu\right)}{RR\left(r,\mu\right)}
\end{equation}
where $DD$, $DR$, and $RR$ are the number of galaxy-galaxy, galaxy-random, and random-random pairs separated by $r$ and $\mu$. 
The random points are given by a set of randomly distributed points with the same selection function as our survey data.

Each object is also given a FKP weight \citep{FKP94} given by:
\begin{equation}
w_i = \frac{1}{1+\bar{n}\left(z_i\right)P\left(k_0\right)}
\end{equation}
where $\bar{n}\left(z_i\right)$ is the number density at the redshift of the $i$th object and $P\left(k_0\right)=20000h^{-3}$ Mpc$^3$ is the 
estimated power spectrum at the BAO scale. 

In addition to the FKP weight, each object is also assigned a systematic weight in order to account for various systematic errors in the data collection. The total systematic weight for an object, $w_{tot}$, is defined in \cite{And13} as:
\begin{equation}
\label{eqn:weights}
w_{tot} = \left(w_{cp} + w_{zf} - 1\right) w_{star} w_{see}
\end{equation}
Here, $w_{zf}$ accounts for failures in acquiring the redshift of the nearest object and $w_{cp}$ accounts for failures in acquiring redshift as a result of being in a close pair with a neighboring object. Further, $w_{star}$ and $w_{see}$ account for systematic relationships between the number density of observed galaxies, stellar densities, and seeing of data.

Once we have computed $\xi_0$ and $\xi_2$ as described in the preceding section, we compute $\omega_0\left(r_c\right)$ and $\omega_2\left(r_c\right)$ by applying Eq.~\ref{eqn:wLdef} using the filter defined in Eq.~\ref{eqn:filter}.
The integrals are computed over the fitting range and evaluated at the same $r_c$'s as the $r$ bins in $\xi$. 

\subsection{Fitting Methodology}
\label{ssec:fitting}

Our multipoles analysis closely follows the analysis conducted in \cite{And14} and \cite{Xu12}. 
We fit both the monopole and quadrupole simultaneously to our model for four nonlinear  parameters: $\alpha$, $\epsilon$, $log\left(B_0^2\right)$, and $\beta$.
This model is constructed based on our expected observations given in Eq~\ref{eqn:expectedxis}:

\begin{equation}
\xi_{0,m}\left(r\right)=B_0^2\xi_{0,t}\left(r\right) + A_0\left(r\right)
\end{equation}
\begin{equation}
\xi_{2,m}\left(r\right) = \xi_{0,t}\left(r\right) + A_2\left(r\right)
\end{equation}
where the subscript $t$ denotes the templates given in Eq~\ref{eqn:expectedxis} and the subscript $m$ denotes our model. 

Our models for $\wL$ are derived from the templates for $\xi_{\ell}$ by applying Eq.~\ref{eqn:wLdef} with the filter defined in Eq.~\ref{eqn:filter} to the expected observations for $\xi_{\ell}$ given in Eq.~\ref{eqn:expectedxis}. 
In short, $\wL$ is a filtered integral of $\xi$. 
As in the multipoles fitting, when fitting $\wL$, we fit $\omega_0\left(r_c\right)$ and $\omega_2\left(r_c\right)$ simultaneously with the same nonlinear parameters.

We detail here the fitted parameters. 
$B_0^2$ is a term that scales the monopole template.
Before fitting, we estimate $B_0^2$ from the offset between our model and the measured correlation function at $r=50h^{-1}$Mpc.
This estimation is then used to normalize the monopole to ensure that $B_0^2 \sim 1$. 
To prevent unphysical negative values, we vary $log\left(B_0^2\right)$ instead of $B_0^2$.
We also apply a Gaussian prior centered at 0 with a standard deviation of 0.4 to keep $B_0^2$ near 1. 

In order to improve the fit, we can choose to add polynomial terms  $A_0(r)$ and $A_2(r)$  to $\omega_0\left(r_c\right)$ and $\omega_2\left(r_c\right)$ when fitting $\wL$, similar to what is done  when we fit the multipoles above. 
$A_l(r)$ is defined by:
\begin{equation}
\label{eqn:nuisparams}
A_l\left(r\right) = \frac{a_{l,1}}{r^2} + \frac{a_{l,2}}{r} + a_{l,3}
\end{equation}
These polynomials are made up of linear nuisance terms used to marginalize out broadband effects like scale-dependent bias and our mismatch between the theoretical redshift-space distorted biased galaxy correlation function and the true correlation function \citep{Xu12}. 
These polynomial terms were used when working with the multipoles. 
However, they were not used with the $\wL$ statistic since we expect that polynomial terms are not needed when fitting $\wL$. 

Our templates are based on the model for the correlation function described in Sec.~\ref{ssec:modeling} with the streaming scale in Eq.~\ref{eqn:FOG} set to $\Sigma_s = 1.4h^{-1}$Mpc. 
The non-linear damping in Eq.~\ref{eqn:2dPS} is set to $\Sigma_{\perp}=6h^{-1}$Mpc and $\Sigma_{||} = 11h^{-1}$Mpc. 
We also allow $\beta$ to vary to better match the amplitude of the quadrupole. 
For $\beta$, we apply a prior centered at 0.4 with a standard deviation of 0.2 to prevent unphysical values. 
Furthermore, since the $\epsilon$ distribution is nearly Gaussian \citep{Xu12}, we apply a 10\% Gaussian prior on $1+\epsilon$ centered on zero to prevent unrealistic values for $\epsilon$. 

The non-linear parameters are fit using a simplex algorithm which steps through the parameter space until the $\chi^2$:
\begin{equation}
\chi^2 = \left(\vec{m}-\vec{d}\right)^TC^{-1}\left(\vec{m}-\vec{d}\right)
\end{equation} 
is minimized. 
Here, $\vec{m}$ is a column vector containing the model, $\vec{d}$ is a column vector containing the data and $C$ is the covariance matrix defined in Sec.~\ref{ssec:covm}.

\begin{table}
\centering
\begin{tabular}{c | c c}
\hline
 $r_{min}$ ($h^{-1}$Mpc) & $\alpha$ & $\epsilon$\\
 \hline
 $22$ & $1.008 \pm 0.028$ & $0.032 \pm 0.022$\\
 $30$ & $0.998 \pm 0.014$ & $0.005 \pm 0.015$\\
 $38$ & $1.001 \pm 0.015$ & $0.000 \pm 0.017$\\
 $46$ & $1.003 \pm 0.015$ & $0.001 \pm 0.022$\\
 $54$ & $1.001 \pm 0.018$ & $0.000 \pm 0.023$\\
 $62$ & $1.000 \pm 0.017$ & $0.000 \pm 0.022$\\
 $70$ & $1.007 \pm 0.026$ & $-0.006 \pm 0.040$\\
 $78$ & $0.991 \pm 0.013$ & $0.002 \pm 0.028$\\
 $86$ & $1.062 \pm 0.030$ & $0.009 \pm 0.083$\\
\end{tabular}
\caption{Average and standard deviation for $\alpha$ and $\epsilon$ for our 600 mocks as $r_{min}$ was varied. They were fit using $\wL$ without polynomial terms.}\label{tab:wLrmin}
\end{table}

When we fit the multipoles, we adopt a fitting range of $46h^{-1}$Mpc to $200h^{-1}$Mpc, which matches the range adopted in \cite{And14}. 
For the $\wL$ fits, we analyzed the possible fitting ranges by fitting our 600 mocks with varying $r_{min}$'s and a fixed $r_{max} = 200h^{-1}$Mpc for our fitting range and recording the best fit $\alpha$ and $\epsilon$ values we obtained. 
Based on the results in Table~\ref{tab:wLrmin}, we concluded that an $r_{min}$ of $46h^{-1}$Mpc would be the best for maintaining a close comparison between the two methods.

\subsection{Error Estimation}
\label{ssec:errorest}
Following the error analysis conducted in \cite{Xu12}, the errors on both the $\wL$ and multipoles fits to data are estimated by computing the probability distribution, $p\left(\alpha,\epsilon\right)$, on a grid of $\left(\alpha,\epsilon\right)$.
For each $\alpha$ and $\epsilon$ in our grid, we fix $\alpha$ and $\epsilon$ and fit the remainder of the parameters using $\chi^2$ as a goodness of fit indicator.
Under the assumption that the likelihood is Gaussian:
\begin{equation}
p\left(\alpha,\epsilon\right) \propto exp\left(\frac{-\chi\left(\alpha,\epsilon\right)^2}{2}\right)
\end{equation}
and normalizing accordingly, we find that:
\begin{equation}
p\left(\alpha\right)=\int p\left(\alpha,\epsilon\right) d\epsilon
\end{equation}
\begin{equation}
p\left(\epsilon\right)=\int p\left(\alpha,\epsilon\right) d\alpha
\end{equation}
Since we assume Gaussian likelihoods, we can take the widths of the distribution $\sigma_\alpha$ and $\sigma_\epsilon$:
\begin{equation}
\sigma_\alpha^2 = \int p\left(\alpha\right)\left(\alpha-\langle\alpha\rangle\right)^2d\alpha
\end{equation}
\begin{equation}
\sigma_\epsilon^2 = \int p\left(\epsilon\right)\left(\epsilon-\langle\epsilon\rangle\right)^2d\epsilon
\end{equation}
as our error estimates.
Here, $\langle x\rangle$ is the expected value of the distribution:
\begin{equation}
\langle x\rangle = \int p\left(x\right)xdx
\end{equation}

For our error estimation, we range $\alpha$ from 0.7 to 1.3 inclusive with a spacing of 0.005 and $\epsilon$ from -0.3 and 0.3 inclusive with a spacing of 0.01.
This yields 121 $\alpha$ bins and 61 $\epsilon$ bins.

%% file: results.tex
\subsection{Testing Mocks}
\label{ssec:testmocks}
%2.1

To appraise the performance of fitting the $\wL$ statistic, we apply the two clustering estimators to the 600 PTHalos mocks and compare the results.
Since the cosmology of the mocks is known, we apply it as the fiducial cosmology and use it in our fitting, we expect best fit values for $\alpha$ near 1 and best fit values for $\epsilon$ near 0.
The average and standard deviation for the fit parameters are shown in Table~\ref{tab:aemocks}.
These results suggest that both methods give consistent values for $\alpha$ and $\epsilon$ with discrepancies within our error bars.
We note that the smaller errors returned by using the $\wL$ statistic are reflected in the $\chi^2$ distribution from fitting the mocks.
The $\chi^2$ distribution produced by fitting $\wL$ yields a tighter distribution, centered closer to 1.

\begin{table}
\centering
\begin{tabular}{c | c c}
\hline
 Method & $\alpha$ & $\epsilon$\\
 \hline
 $\wL$ & $1.003 \pm 0.015$ & $0.001 \pm 0.017$\\
 Multipoles & $1.004 \pm 0.016$ & $0.002 \pm 0.018$\\
 \hline
\end{tabular}
\caption{Average and standard deviation for $\alpha$ and $\epsilon$ for our 600 mocks, fitted using $\wL$ without polynomial terms and the multipoles method.}\label{tab:aemocks}
\end{table}

\subsection{Fitting Data}
\label{ssec:fitdata}

We also test the performance of the two methods on data from DR11.
Our results are summarized in Table~\ref{tab:aedata}.
\begin{table}
\centering
\begin{tabular}{c | c c}
\hline
Method & $\alpha$ & $\epsilon$\\
 \hline
$\wL$ & $1.029 \pm 0.020$ & $-0.003 \pm 0.023$\\
Multipoles & $1.025 \pm 0.014$ & $-0.010 \pm 0.019$\\
Anderson, et al. (2014b) & $1.025 \pm 0.014$ & $-0.010 \pm 0.019$\\
 \hline
\end{tabular}
\caption{Best fit values and standard deviation for $\alpha$ and $\epsilon$ for DR11 data, fitted using $\wL$ and the multipoles method. We also include as a reference the fit results for multipoles with the de-wiggled template conducted in Anderson et al. (2014b).}\label{tab:aedata}
\end{table}
\begin{figure}
\includegraphics[width=\linewidth]{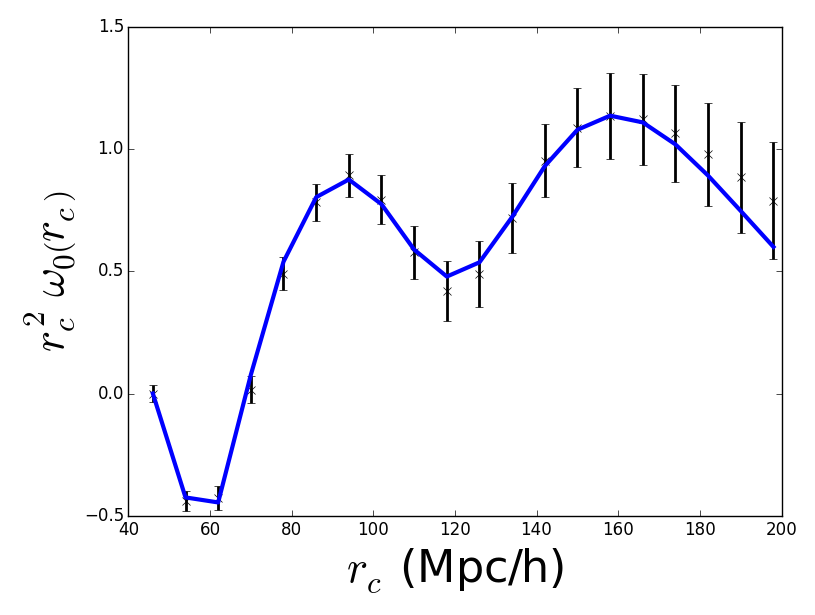}
\includegraphics[width=\linewidth]{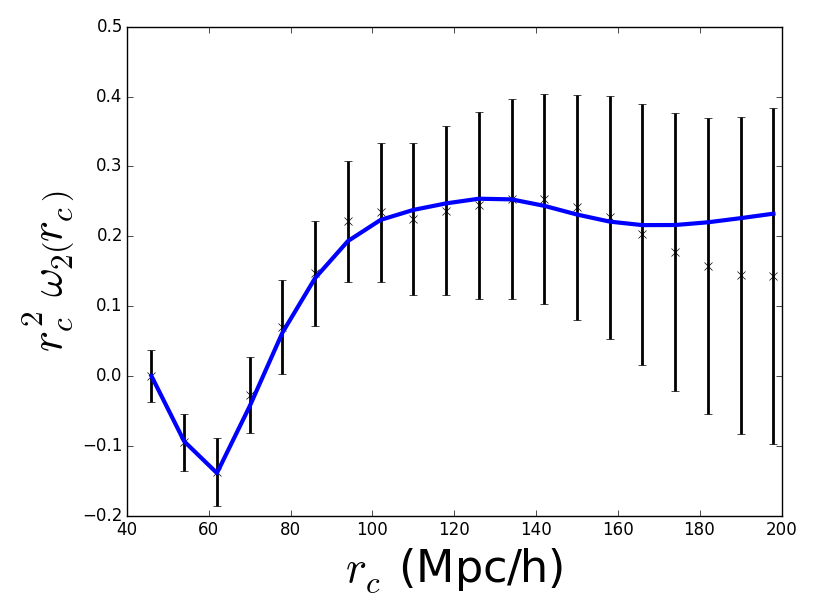}
\caption{Fits to DR11 data using $\wL$. $\ell=0$ is above and $\ell=2$ is below. The crosses represent the DR11 data and the blue line is our best fit.}\label{fig:wLdatafit}
\end{figure}
Here, the errors on our parameters are computed using the analysis outlined in~\ref{ssec:errorest}.
Figure~\ref{fig:wLdatafit} depicts our fit to the $\wL$ statistic for DR11 data. 

Our results for $\alpha$ are in agreement for the two methods tested here.
The results we obtained for $\epsilon$ also agree within 1 sigma.
These numbers suggest that the two methods are consistent with each other.
The results of our multipoles fitting are also in agreement with the numbers given in \cite{And14}, which is reassuring.

%% file: advantages.tex
In this section we conduct a brief investigation into a possible advantage of using the $\wL$ statistic.
Based on \cite{Xu10}, we expect $\wL$ to respond better to large scale abnormalities than the multipoles method.
In order to test this response we conduct two tests comparing the performance of the $\wL$ method against the multipoles method.  

\subsection{Polynomial Addition of Extra Power}
\label{ssec:polyadd}
Our first test consists of directly adding a polynomial to $\xi_0\left(r\right)$ to introduce abnormalities at large scales.
The choice of polynomial is inspired in part by form of large scale power we observed in early BOSS DR9 data as well as in the quasar correlation function. 
The polynomial added, $p\left(r\right)$, is of the following form:
\begin{equation}
p\left(r\right) = \frac{1}{r^2}\left(2x^3\right)^2\left(1-x^3\right)^2\left(\frac{1}{2}-x^3\right)\left(\frac{S}{0.02}\right)
\end{equation}
Here, $x = r / 275$ and $S$ is a scale factor we adjust to change the amplitude.
\begin{figure}
\includegraphics[width=\linewidth]{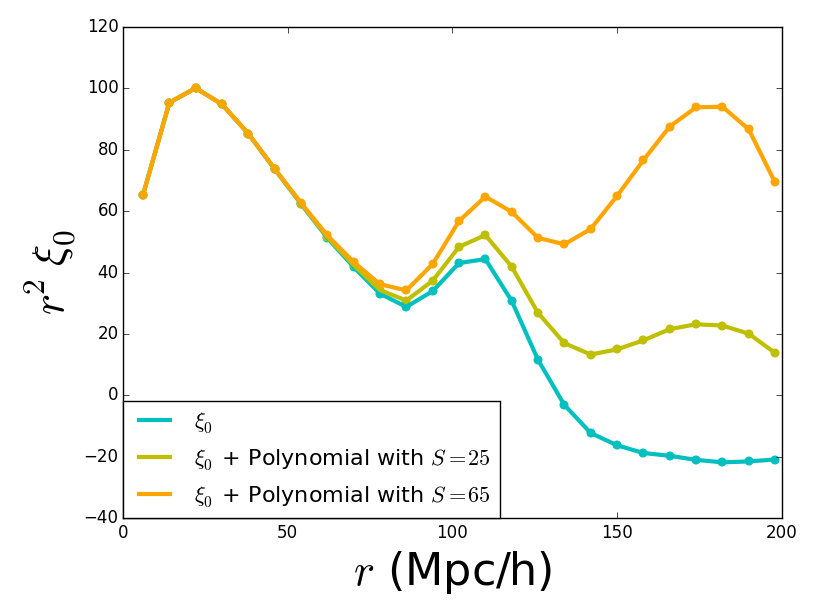}
\caption{An unmodified $\xi_0$ compared to 2 modified $\xi_0$s.}\label{fig:Xiaddpoly}
\end{figure}
Figure~\ref{fig:Xiaddpoly} depicts the addition of polynomials with varying $S$ to $\xi_0$. 
We apply the above treatment to the correlation functions for our 600 mocks and repeat the same fitting analysis applied earlier (Sec.~\ref{sec:analysis}).
\begin{table}
\centering
\begin{tabular}{c | c c}
\hline
 Method & $\alpha$ & $\epsilon$\\
 \hline
  Mult. $S=0$ & $1.004 \pm 0.016$ & $0.002 \pm 0.018$\\
 Mult. $S=25$ & $1.004 \pm 0.016$ & $0.001 \pm 0.018$\\
 Mult. $S=45$ & $1.004 \pm 0.017$ & $-0.000 \pm 0.020$\\
 Mult. $S=65$ & $1.004 \pm 0.017$ & $-0.001 \pm 0.020$\\
 \hline
 $\wL, S=0$ & $1.003 \pm 0.015$ & $0.001 \pm 0.017$\\
 $\wL, S=25$ & $1.003 \pm 0.015$ & $0.000 \pm 0.018$\\
 $\wL, S=45$ & $1.003 \pm 0.016$ & $0.000 \pm 0.018$\\
 $\wL, S=65$ & $1.003 \pm 0.016$ & $-0.001 \pm 0.019$\\
 \hline
\end{tabular}
\caption{Average and standard deviation for $\alpha$ and $\epsilon$ for our 600 mocks, fit using $\wL$ (without polynomial terms) and the multipoles method with varying values of $S$.}\label{tab:aeaddpoly}
\end{table}

%\newpage
The best fit values for various scale factors $S$ are presented in Table~\ref{tab:aeaddpoly}. 
Again, since we know the fiducial cosmology is set to that of the mocks, we expect $\alpha$ to be approximately 1 and $\epsilon$ to be approximately 0.

We see that, as the value of $S$ is increased, the average best fit values and standard deviations for the multipoles fit remain relatively unchanged. 
The average best fit values and standard deviations for the $\wL$ fit also remain relatively unchanged from the default of $S=0$. 
The $\wL$ fit performs similar to the multipoles fit in response to changes in $S$.

Our results indicate that even without polynomial nuisance parameters in the $\wL$ fit, it manages to match the performance of the multipoles fit [when extra large scale power is added], which uses polynomial nuisance parameters.
 
\subsection{Weight Application}
\label{ssec:weights}
\begin{figure}
\includegraphics[width=\linewidth]{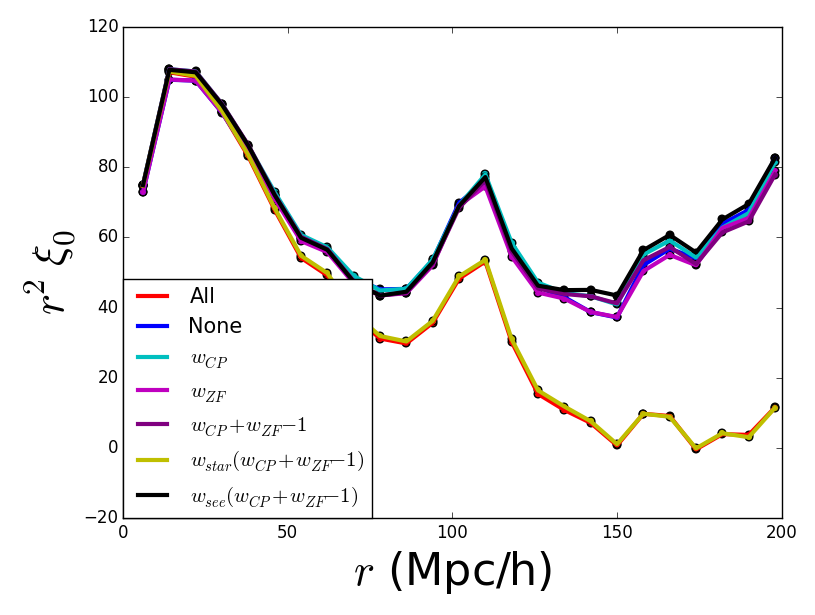}
\includegraphics[width=\linewidth]{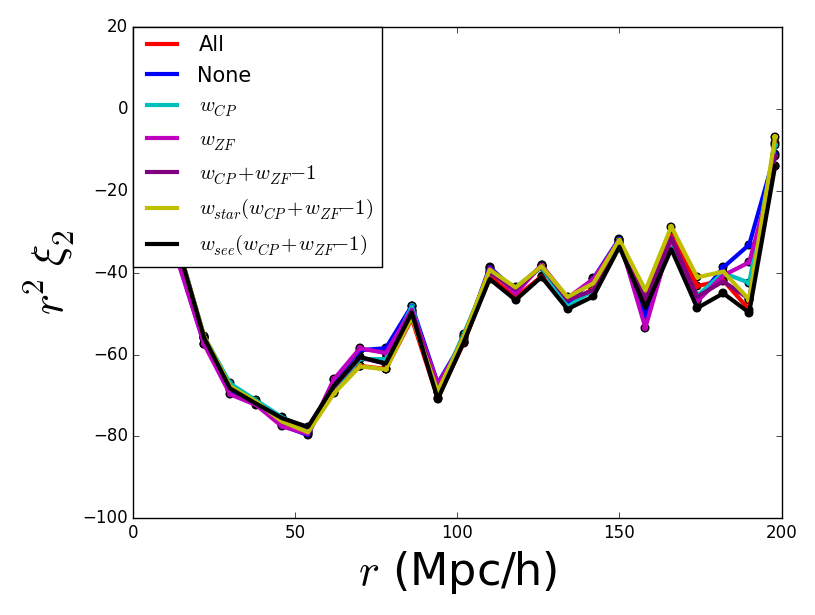}
\caption{The effects of different weight applications on $\xi$, computed from DR11 data. The top plot depicts $\xi_0$ and the bottom plot depicts $\xi_2$. The top plot depicts the effects of different weight application on $\xi_0\left(r\right)$.}\label{fig:Xiweights}
\end{figure}

\begin{figure}
\includegraphics[width=\linewidth]{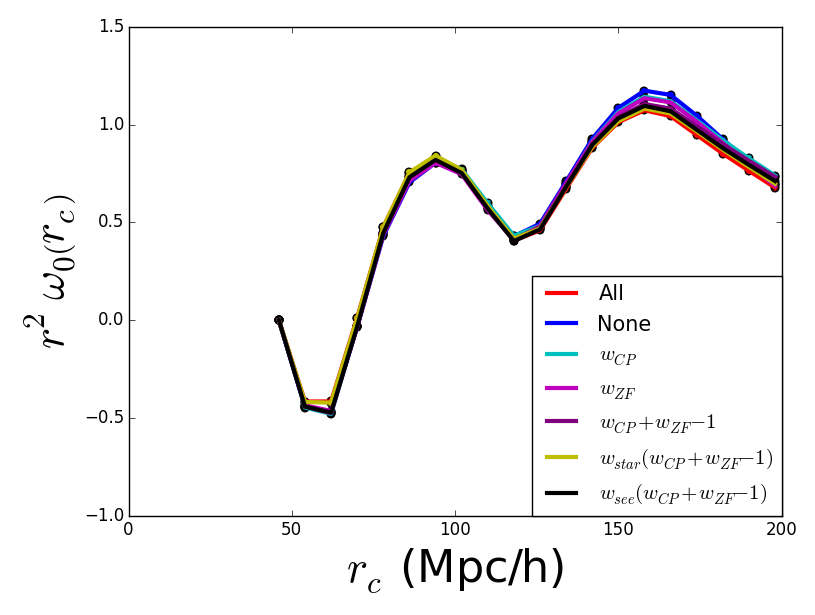}
\includegraphics[width=\linewidth]{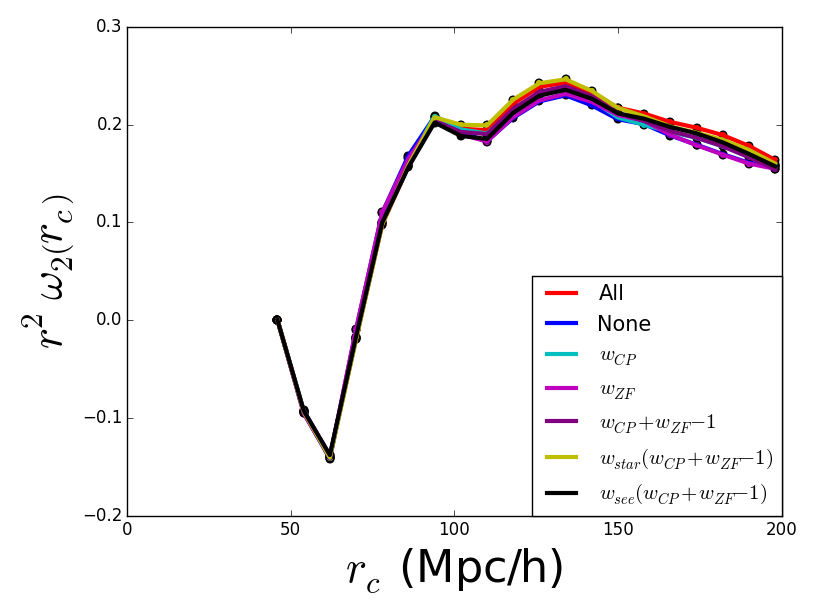}
\caption{The effects of different weight applications on $\wL$, computed from DR11 data. The top plot depicts $\omega_{0}\left(r_c\right)$ and the bottom plot depicts $\omega_{2}\left(r_c\right)$. We see that the application of various systematic weights do not affect the shape of the $\wL$ statistic, making it a stable estimator in the face of systematics.}\label{fig:wLweights}
\end{figure}

Our second test investigated how applying subsets of the weights ($w_{zf}, w_{cp}, w_{star}, w_{see}$) when computing the correlation function for the DR11 data affected the performance of each fitting method. 
The SDSS-III BOSS analysis derived these weights in an effort to reduce the effect of observational systematics on data analysis. 
However, there are no errors computed for these weights.
It would be advantageous if the $\wL$ method conducted without weights returned similar results to the $\wL$ method conducted with all the weights, making the absence of errors in the weight computation less of a problem. 
The different weight applications, given by different forms of Eq.~\ref{eqn:weights} and their effect on $\xi_{\ell}\left(r\right)$ and $\omega_{\ell}\left(r_c\right)$, are depicted in Figures~\ref{fig:Xiweights} and~\ref{fig:wLweights}.

\begin{table}
\centering
\begin{tabular}{c | c c c c}
\hline
 Method and Weights & $\Delta\alpha$ & $\Delta\epsilon$ \\
 \hline
 Mult. All & 0.0000 & 0.0000\\
 Mult. None & 0.0027 & 0.0002\\
 Mult. $w_{CP}$ & 0.0043 & 0.0016\\
 Mult. $w_{ZF}$ & 0.0003 & 0.0004\\
 Mult. $(w_{CP} + w_{ZF} - 1)$ & 0.0011 & 0.0006\\
 Mult. $w_{star}(w_{CP} + w_{ZF} - 1)$ & 0.0004 & 0.0003\\
 Mult. $w_{see}(w_{CP} + w_{ZF} - 1)$ & 0.0006 & 0.0009\\
 \hline
 $\wL$; All & 0.0000 & 0.0000\\
 $\wL$; None & 0.0037 & 0.0024\\
 $\wL$; $w_{CP}$ & 0.0066 & 0.0012\\
 $\wL$; $w_{ZF}$ & 0.0004 & 0.0019\\
 $\wL$; $(w_{CP} + w_{ZF} - 1)$ & 0.0026 & 0.0015\\
 $\wL$; $w_{star}(w_{CP} + w_{ZF} - 1)$ & 0.0012 & 0.0005\\
 $\wL$; $w_{see}(w_{CP} + w_{ZF} - 1)$ & 0.0014 & 0.0009\\
 \hline
\end{tabular}
\caption{Absolute difference of best fit $\alpha$ and $\epsilon$ for each weight application from the default of applying all weights.}\label{tab:weightfitres}
\end{table}

Table~\ref{tab:weightfitres} contains the results of our tests. 
We find that in the majority of weight applications, although the multipoles fit performs slightly better, the two fitting methods remain comparable.
%The best fit values for the two methods when using a modified weight application change at a similar level from the default of applying all the weights.
%In some cases, fitting the multipoles is more stable while in other cases, fitting the $\wL$ statistic is more stable.
As in the previous test, our results indicate that even without polynomial nuisance parameters in the $\wL$ fit, the $\wL$ fit nearly matches the performance of the multipoles fit in the face of systematics even though the multipoles fit uses polynomial nuisance parameters.

%% file: conclusion.tex
Observational systematics in cosmological surveys can impart systematic fluctuations at large scales.
It is useful to investigate new statistics that are less sensitive to these large scale fluctuations. 
In this paper, we consider one such statistic, $\wL$. 
In order to determine the power of the $\wL$ statistic compared to the standard multipoles analysis, we applied both methods to fit for $\alpha$ and $\epsilon$ in 600 PTHaloes mocks and data from SDSS DR11.
After comparing the best fit values and standard deviations returned by each method, we conclude that fitting with the $\wL$ statistic without polynomial nuisance terms is comparable to the standard multipoles fitting. 

In addition, we also investigated the performance of each method in response to abnormalities in the correlation function at large scales. 
In line with the predictions in \cite{Xu10} we find that fitting the $\wL$ statistic is less sensitive to abnormalities generated by the addition of extra power (in the form of a polynomial) at large scales than the standard multipoles fit.
The $\wL$ statistic matches the performance of fitting the multipoles even without the use of polynomial nuisance parameters.
We observe similar results when analyzing the response of the $\wL$ statistic to changes in the application of systematic weights to the data.
Without using polynomial nuisance parameters, the $\wL$ statistic returns similar results as those given by fitting the multipoles.

The investigation here suggests that the $\wL$ statistic could be advantageous compared to the standard multipoles method when dealing with abnormalities at large scales.
Fitting the $\wL$ statistic eliminates the need for the polynomial nuisance parameters which must be used in conventional methods for fitting the multipoles.

%% file: acknowledgement.tex
This work is partially supported by NASA NNH12ZDA001N-EUCLID. 
S.H. and K.O. are partially supported by DOE-ASC, NASA and the NSF.

Numerical computations for the PTHalos mocks were done on the Sciama
High Performance Compute (HPC) cluster which is supported by the ICG,
SEPNet and the University of Portsmouth.

Funding for SDSS-III has been provided by the Alfred P. Sloan Foundation, the Participating
Institutions, the National Science Foundation, and the U.S. Department of Energy.

SDSS-III is managed by the Astrophysical Research Consortium for the
Participating Institutions of the SDSS-III Collaboration including the
University of Arizona,
the Brazilian Participation Group,
Brookhaven National Laboratory,
University of Cambridge,
Carnegie Mellon University,
University of Florida,
the French Participation Group,
the German Participation Group,
Harvard University,
the Instituto de Astrofisica de Canarias,
the Michigan State/Notre Dame/JINA Participation Group,
Johns Hopkins University,
Lawrence Berkeley National Laboratory,
Max Planck Institute for Astrophysics,
Max Planck Institute for Extraterrestrial Physics,
New Mexico State University,
New York University,
Ohio State University,
Pennsylvania State University,
University of Portsmouth,
Princeton University,
the Spanish Participation Group,
University of Tokyo,
University of Utah,
Vanderbilt University,
University of Virginia,
University of Washington,
and Yale University.

We would like to thank Tommy Dessup and Xiaoying Xu for assistance and inspiration provided during the course of this work.
We would also like to thank Martin White and his insightful comments and advice.